\documentclass[12pt]{article}
\usepackage{graphicx}
\usepackage[english]{babel}

\begin{document}

\title{Quantum and semiclassical spin networks: from atomic and
       molecular physics to quantum computing and gravity}

\author{Vincenzo Aquilanti$^{\dag}$, Ana Carla P. Bitencourt$^{\dag}$,\\
        Cristiane da S. Ferreira$^{\dag}$,\\
        Annalisa Marzuoli$^{\ddagger}$ and Mirco Ragni$^{\dag\,(1)}$}

\maketitle

{\small\noindent $^{\dag}$Dipartimento di Chimica, Universit\`a di Perugia,
        via Elce di Sotto 8, 06123 Perugia (Italy)\\
        $^{\ddagger}$Dipartimento di Fisica Nucleare e Teorica,
        Universit\`a degli Studi di Pavia and Istituto Nazionale
        di Fisica Nucleare, Sezione di Pavia, via A. Bassi 6,
        27100 Pavia (Italy)

\begin{abstract}
The mathematical apparatus of quantum--mechanical angular momentum
(re)coupling, developed originally to describe spectroscopic
phenomena in atomic, molecular, optical and nuclear physics, is
embedded in modern algebraic settings which emphasize the
underlying combinatorial aspects. $SU(2)$ recoupling theory,
involving Wigner's $3nj$ symbols, as well as the related problems
of their calculations,  general properties,  asymptotic limits for
large entries, plays nowadays a prominent role also in quantum
gravity and quantum computing applications. We refer to the
ingredients of this theory --and of its extension to other Lie and
quantum group-- by using the collective term of `spin networks'.

Recent progress is recorded about the already established connections with the
mathematical  theory of discrete orthogonal polynomials
(the so--called Askey Scheme),
providing powerful tools based on asymptotic expansions, which
correspond on the physical side to various levels of semi--classical
limits. These results are useful not only in theoretical molecular
physics but also in motivating algorithms for the computationally
demanding problems of molecular dynamics and chemical reaction theory,
where large angular momenta are typically involved.
As for quantum chemistry, applications of these techniques include
selection
and classification of complete orthogonal basis sets in atomic and
molecular problems, either in configuration space (Sturmian orbitals)
or in momentum space.
In this paper we list and discuss some aspects of these developments
--such as for instance the hyperquantization algorithm-- as well as a few
applications to quantum gravity and topology, thus providing evidence of
a unifying background structure.
\end{abstract}

\section{Introduction}
The (re)coupling theory of many $SU(2)$ angular momenta --the
Racah-Wigner tensor algebra-- is the most exhaustive framework
in dealing  with interacting many--angular momenta quantum systems
that can be modeled by means of conservative polylocal two--body
interactions \cite{BILO9}. The essential features of this algebra can be encoded,
for each fixed number $N=(n+1)$  of
angular momentum variables, into a combinatorial object, the {\em
spin network graph}. Vertices are associated with
finite--dimensional, binary coupled Hilbert spaces while edges
correspond to either phase  or Racah transforms (implemented by
$6j$ symbols) acting on states in such a way that the quantum
transition amplitude between any pair of vertices is provided
by a suitable $3nj$ symbol.

One of the goals of this paper is to promote this combinatorial setting
to (families of)  {\em computational quantum graphs} (Sec.\ref{sec:due}),
to be thought of as a sort of `abacus' encoding diagrammatical rules
encountered in quantum collision theory \cite{254} or, more in general,
as computational spaces of simulators able to support quantum algorithms
for computational problems arising in mathematics and theoretical physics
\cite{MARA2}. Progress in the semiclassical limits is accounted for in
Sec. \ref{sec:tre}. A few examples are then briefly addressed, emphasis
being placed on those semiclassical versions of the construction, which turn
out to play a crucial role in many different contexts, ranging from molecular physics
(Sec. \ref{sec:quattro}) and quantum computing and discrete quantum gravity models
(Sec. \ref{sec:cinque}). An outlook to applications and a list of perspectives for future
work are collected in the concluding Sec. \ref{sec:sei}.

\section{\label{sec:due}$SU(2)$ recoupling schemes as spin network graphs}

Important ingredients of modern theoretical physics and chemistry
are intimately linked to the mathematical theory of classical
(as opposed to $q$--deformed) orthogonal polynomials, in particular
of  hypergeometric families. The recently proposed formalization of
the latter within the Askey and Nikiforov schemes \cite{KOE98,NIK91}
places Racah polynomials --and their $q$--deformed analogs \cite{BIE95}--
at the  top of a hierarchy from which all the most relevant families
are obtained as particular cases or via suitable limiting procedures.
Although the study of such morphology may be considered as an exercise
in special function theory, relationships among families, addition formulas,
linearization formulas and sum rules look  like obscure manipulations
of abstract quantities unless  a coherent interpretation arising from
physical applications could be disclosed. Physical applications, on the
other hand, can lead to new insights into previously unnoticed properties
and relationships.

Dealing with the quantum theory of angular momentum, Racah found a finite
sum expression for the basic `recoupling' coefficients or $6j$ symbols,
representing matrix elements for the orthogonal basis transformation between
two alternative binary coupling schemes of three angular momenta.
Extensive developments of graphical techniques, initiated by the Yutsis
school \cite{YUT62}, allow one to deal efficiently with recouplings of
$(n+1)$ angular momenta, involving the so--called $3nj$ symbols (see the
handbook \cite{VAR88} and the review \cite{BILO9} (Topic 12) also for a
complete list of references).

An introduction and some nomenclature are exhibited in \cite{254}, where
the `spin networks', as well as the underlying `moves', are shown to be
closely related to the tree--like graphical representation of hyperspherical
coordinates and harmonics, as originally suggested in Ref. \cite{VIL66:645}.

Explicit formulas and  basic relationships among recoupling
coefficients and harmonic superpositions (or
{\em transplant} coefficients) and Racah polynomials can be found
in \cite{253} and in references therein. The close relationship of
both vector recoupling coefficients and superposition matrix
elements between alternative hyperspherical harmonics with
orthogonal polynomials of a discrete variable made it possible to
develop a common classification scheme, from which a number of
relations and properties can be derived \cite{254}. The important
case of $S^3$ harmonics, discussed in \cite{216} from the viewpoint of
hydrogenoid orbitals in momentum space, is reviewed in \cite{252},
where further applications  for quantum chemistry are outlined.

The underlying mathematical background of the
developments discussed in \cite{254}, involves
objects like the $3nj$ symbols of angular momentum theory (and
their next of kin, the superposition coefficients) which are
candidates for further study as orthogonal polynomials of $(n-1)$
discrete variables.

The basic features of spin network graphs are illustrated in
Fig. \ref{fig1} for the significant case of binary coupling schemes of
four angular momenta and their recoupling represented by
a Wigner $9j$ coefficient. We refer to \cite{254} for
more details on their use as
`abacus', namely as graphical devices for practical calculations.
For the construction of the abacus a proper look in terms of
`presentation' of discrete groups shows that the basic tools
are the three operations corresponding to the presentation of
the icosahedral group \cite{COX65,RAS81:1}.
From a geometrical viewpoint, the abacus essentially mixes
(according to a well--known presentation of the icosahedral
symmetries) pentagonal and hexagonal cycles, whose deep origin
can be traced in abstract algebraic theories\cite{MAJ90:1}.

\begin{figure}
 \centering\includegraphics[width=.8\textwidth]{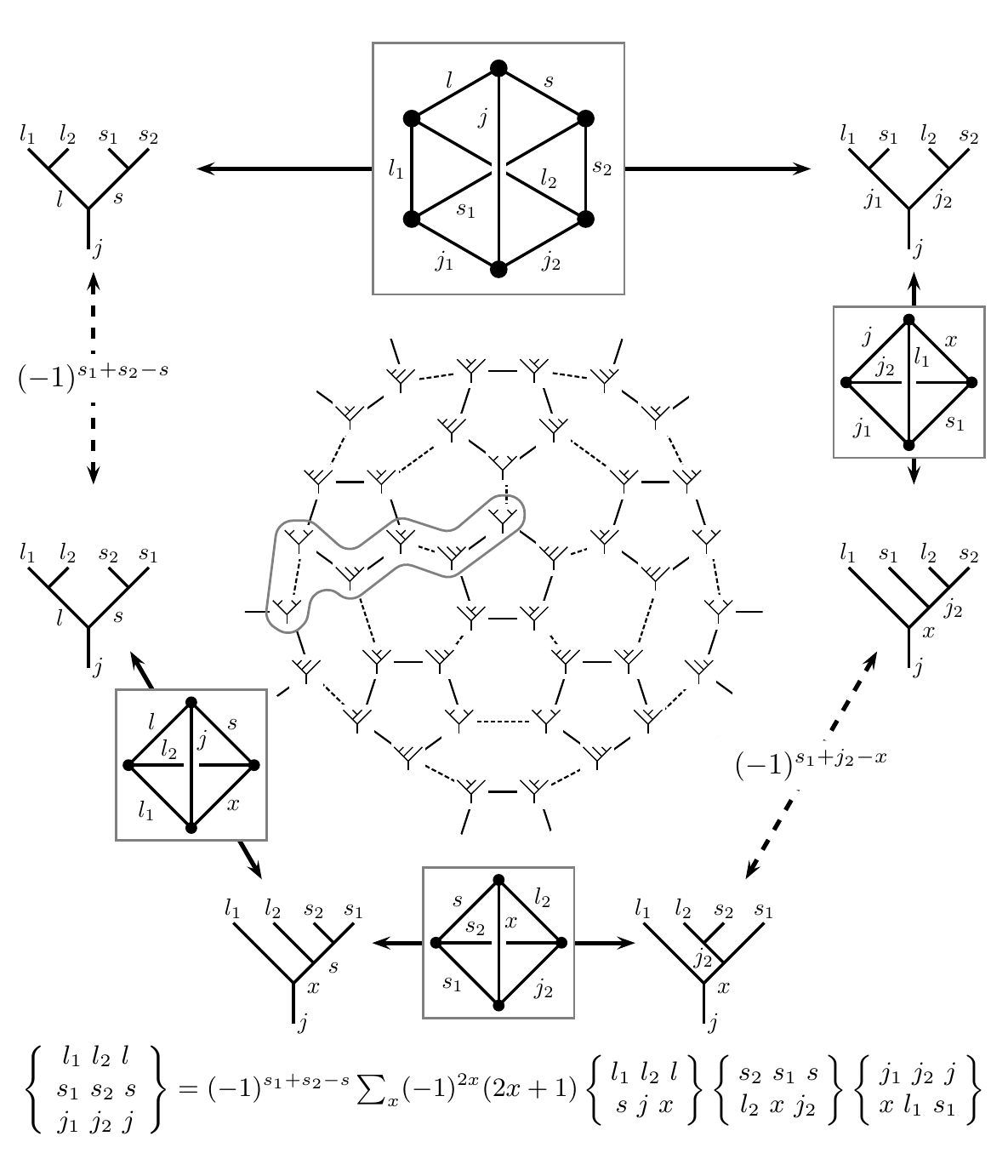}
 \caption{\label{fig1}\linespread{1}\footnotesize{
Spin networks are illustrated here for the case of two quantum systems, 1 and 2, each endowed
with both spin and orbital angular momentum, giving a state with total angular
momentum $j$ (coupling described by spin--orbit terms in the Hamiltonian).
The two alternative ways of pairing $s_1, s_2,l_1,l_2$, traditionally
referred to as $ls$ and $jj$ couplings, are encoded in the top left and right
`binary trees'. Their recoupling --achieved by joining edges with identical labels-- leads to the
$3$--valent, not planar graph (of the Yutsis--type, as referred to in the literature).
The associated recoupling matrix elements give the $9j$ symbol, represented by the
$3\times 3$ array on the left--hand side of the formula. In the center a portion of
what we call an `abacus', namely an aid in obtaining the procedure for getting the decomposition
illustrated in the rest of the external region of the figure: the three complete graphs
on $4$ vertices are associated with suitably labelled $6j$ symbols, while dotted arrows
correspond to phase changes. The structure of the spin network graph displays continuous
and dashed lines as well as typical pentagonal and hexagonal cycles (providing an overall
`fullerene'-like pattern), associated with the Biedenharn--Elliott and Racah identities,
respectively.}}
\end{figure}

\section{\label{sec:tre}Semiclassical spin networks}

According to  Bohr correspondence principle, classical concepts become increasingly
valid in  regimes where all (or just a few) quantum numbers are `large'
(as will be illustrated in Sec. \ref{sec:quattro}, such regimes are quite commonly
encountered in every--day analysis of atomic and molecular dynamical processes).

In handling with angular momenta variables measured in units of
$\hbar$, the classical limit $\hbar \rightarrow 0$ implies in particular that,
for finite angular momenta, the $j$--quantum numbers
are much larger than one.

In the case in which  all the six angular momenta are large,
the Racah $6j$ symbol admits a well defined {\em
asymptotic limit}, whose absolute square (probability)
corresponds to the {\em classical limit} of the associated
physical observables.
Following \cite{PORE,SCGO1,SCGO2} and
\cite{BILO9} (Topic 9) (where a self contained treatment of
various asymptotics of angular momentum functions is given,
together with the list of original references), recall
that the square of the symbol has the
limiting value given by the Wigner formula
\begin{equation}\label{6jsquare}
\left\{\begin{array}{ccc}
a & b & d\\
c & f & e
\end{array}\right\}^{\,2}\;\;\sim\;\; \frac{1}{12 \pi\,V}
\end{equation}
where $V$ is the Euclidean volume of the tetrahedron formed by
the six angular momentum `vectors'.

A major breakthrough in semiclassical analysis is provided by  the
Ponzano--Regge asymptotic formula for the $6j$ symbol \cite{PORE}
\begin{equation}\label{PRasymt}
\left\{\begin{array}{ccc}
a & b & d\\
c & f & e
\end{array}\right\}
\;\sim\;\; \frac{1}{\sqrt{24 \pi V}}\;
\exp\,\left\{i\,\left(\sum_{r=1}^{6}\,\ell_r \, \theta_r
\,+\,\frac{\pi}{4} \right)\right\}\;,
\end{equation}
\noindent where the limit is taken for all entries $\gg 1$ (recall
that $\hbar =1$) and $\ell_r \equiv j_r +1/2$ with
$\{j_r\}=\{a,b,c,d,e,f\}$. $V$ is now the Euclidean volume of the
tetrahedron with edges of lengths $\{\ell_r\}$, calculated by
using the Cayley determinant (note the shift $j \rightarrow j+
1/2$ with respect to the variables employed in calculating the
volume in (\ref{6jsquare})) and finally $\theta_r$ is the angle
between the outer normals to the faces which share the edge $\ell_r$.

The probability amplitude (\ref{PRasymt}) has the form of a
semiclassical wave function since the factor $1/\sqrt{24 \pi V}$
is slowly varying with respect to the spin variables, while the
exponential is a rapidly oscillating dynamical phase (such
behavior complies with the fact that the square of this asymptotic
expression reproduces Wigner's result (\ref{6jsquare})). Moreover,
according to Feynman path sum interpretation of quantum mechanics,
the argument of the exponential represents a classical action, and
indeed it can be read as $\sum p\,\dot{q}$ for pairs $(p,q)$ of
canonical variables (angular momenta and conjugate angles).
A mathematically sophisticated analysis of Ponzano--Regge
result can be found in \cite{319}, where recent developments
in the study of the asymptotics of $3j$ and $6j$ symbols are
also reviewed.

An interesting issue --arising in connection with the
interpretation of spin networks as computational quantum graphs--
concerns the phenomenon of {\em disentaglement}. One of the main
features of states belonging to the binary coupled Hilbert spaces,
whose `recoupling' gives the $6j$ Racah coefficient, is to
represent effectively `entangled' quantum states (namely states
that cannot be reduced to a product of states containing quantum
numbers of the individual components). Then the Racah transform
(implemented by means of a $6j$ symbol, see the lefthand side of
(\ref{PRasymt})), being a quantum transition amplitude, takes care
of the fact that the operators $\mathbf{J}_{d}^2$ and
$\mathbf{J}_{e}^2$, whose eigenvalues are $d(d+1)$ and $e(e+1)$
respectively, do not belong to the same set of mutually commuting
operators and thus cannot be measured simultaneously. On the other
hand, when reaching the semiclassical limit as can be seen from
the righthand side of (\ref{PRasymt}), the six entries of the $6j$
symbol appear to be on the same footing, a feature that can be
viewed as a `disentanglement' of the underlying `semiclassical'
spin networks. Work is in progess for what concerns the $9j$
symbol \cite{3xx}.

\section{\label{sec:quattro}Molecular Physics}

\subsection{\label{sec:quattroA}Atomic collisions and molecular spectroscopy}
A presentation of angular momentum theory from the viewpoint of these applications,
is given by Zare \cite{ZARE}.
The theory developed about thirty years ago in \cite{70} dealt with
five alternative representations for the quantum-mechanical
close-coupling formulation  \cite{70,71,73} of the motion
along the internuclear distance of a vibrating diatomic molecule
or colliding atoms having internal (spin and/or electronic) angular momenta.
This unified frame transformation approaches of atomic collision
theory and concepts of molecular spectroscopy, is originally due to Hund.
The physical picture and the relevant
nomenclature are reviewed in Ref \cite{200}, see also \cite{245}.
Explicitly \cite{218}, starting from a sum rule equivalent to the
Biedenharn--Elliott relation, (a pentagonal closed path on the abacus of Fig. \ref{fig1})
we obtain the definition of a $6j$ symbol as a sum of four $3j$ symbols by taking a proper
limit, since one angular momentum is much larger than the others\cite{200}.

The coupling schemes of four angular momenta are
illustrated in Ref.\cite{200} as the basic ingredients underlying the
classification of the five Hund cases and the relationships among
representations. It was shown that the transition from one coupling
scheme to another is performed by an orthogonal transformation
whose matrix elements can be written in terms of 6j symbols. In
the pentagonal arrangement of five alternative coupling schemes
for four angular momenta (represented by the tree-like graphs at
the vertices) \cite{218}, connections (the sides of the pentagon)
are realized by orthogonal matrices and are related to $6j$ symbols.
This was the archetype for the abacus in Fig. \ref{fig1}.
Recent extensions of this approach have shown that in a general
theory of interacting open shell atoms, $3nj$ symbols up to $n=6$ occur \cite{KREMS}.

\subsection{\label{sec:quattroB}Reaction dynamics: Hyperquantization}

An important message that we have learned from work reviewed in Sec. \ref{sec:quattroA}
is the view that a continuous variable limit is obtained at high angular
momenta from the discrete structure typical of quantum mechanics.
The opposite viewpoint can be considered as well, namely the semiclassical
limit may describe a continuous structure, and quantum angular momentum
algebra provides discretization.

The search for both alternative reference frames and  angular momentum coupling schemes
has been a major challenge in quantum mechanics since its origin, and tranformations
among them are represented by vector--coupling and recoupling coefficients, respectively.
The relevant equations can then be formulated in terms of quantum numbers,
which approximately correspond to constants of motion of the systems under study.
Fundamental advances have been  achieved over the years within this framework:
in the last Fifties Jacob and Wick introduced the helicity formalism, widely used
for the theoretical treatment of a variety of collisional problems; extending Hund's
introduction of alternative coupling schemes for a diatomic molecule carrying
electronic, spin, and rotational angular momenta (See Sec. \ref{sec:quattroA}).
These developments fit into the frame transformation theory pioneered by Fano
and coworkers in the Seventies.

Indeed, for general anisotropic interaction, discretization
procedures can be introduced by exploiting Racah algebra, which
fosters the introduction of alternative coupling schemes labeled
by `artificial' quantum numbers. This method has been shown to provide an
elegant and powerful tool for the solution of the reactive
scattering Schr\"{o}ndinger equation and, at present, a
considerable number of methods have been developed in this spirit,
among which the `hyperquantization' algorithm, outlined in greater
detail in a number of references \cite{217,218,248}. The technique relies on the
hyperspherical coordinate approach when used for few-body processes,
including rearrangement. For instance, in a reactive triatomic process, the
reaction coordinate is represented by the hyperradius, which is a
measure of the total inertia of the system, and an adiabatic
representation of the total eigenfunction with respect to this
coordinate is adopted. In such a way, a quantization problem on
the surface of the a hypersphere (in this case the sphere in a
$6$--dimensional Euclidean space) must be solved for fixed values of the hyperradius. Then
coupled--channel equations are integrated applying a standard
propagation procedure. The success of this approach is strictly
dependent on the accuracy and the effectiveness of the method used
to solve the fixed hyperradius problem. The computation of the
adiabatic eigenvalues containing detailed information on the
structure, rotations, and internal modes parametrically in the
hyperrradius is typically very demanding. The hyperquantization
algorithm exploits the peculiar properties of the discrete
analogues of hyperspherical harmonics, {\em i.e.} generalized $3j$
symbols or Hahn polynomials,  orthogonal on a grid of points that
span the interaction region \cite{195}. The computationally
advantageous aspect of this algorithm, besides the elegance of
unifying under the language of angular momentum theory the
dynamical treatment of a reaction, is the structure of the
Hamiltonian matrix: its kinetic part is simple, universal, highly
symmetric, and sparse, while the potential displays the diagonal form
characteristic of the stereodirected representations of the
previous section. The hyperquantization algorithm, when
implemented for reactive scattering calculations \cite{218},
allows considerable savings in memory requirements for storage and
in computing time for the building up and diagonalization of large
basis sets, exploiting the sparseness and the symmetry properties
of the Hamiltonian matrix.

\section{\label{sec:cinque}Quantum gravity and quantum computing}
\subsection{Discretized quantum gravity in dimension $3$}

There exists an intriguing physical interpretation of the Ponzano--Regge
asymptotics (\ref{PRasymt})
once we recognize that the expression in the exponential represents the classical
Regge action \cite{REG} --namely the discretized version of Einstein--Hilbert
action of General Relativity -- for the tetrahedron associated with the $6j$ symbol
in the semiclassical limit.

In Fig. \ref{fig1} the tetrahedral symmetry of each
$6j$ symbol is encoded into the corresponding  $4$--vertex graph,
where the six entries of the symbol are associated with edges and
its four triads with faces of a tetrahedron embedded in Euclidean $3$--space
(see {\em e.g.} \cite{VAR88} for more details on the symmetries and the
geometrical properties of  this symbol).

In Regge's approach to General Relativity, the edge lengths of a `triangulated spacetime'
are taken as discrete counterparts of the metric tensor appearing in the usual action
for gravity and angular variables (deficit angles) are related to the scalar
curvature obtained from the Riemann tensor. Technically speaking, a Regge spacetime
is a piecewise linear (PL) manifold of dimension $D$ dissected into
simplices, namely triangles in $D=2$, tetrahedra in $D=3$, 4-simplices in $D=4$
and so on.
`Regge Calculus' became in the early  80's the starting point for a novel approach to
quantization of General Relativity known as simplicial quantum gravity (see
\cite{WIL} and references therein). The quantization procedure most commonly adopted
is the Euclidean path sum approach, namely the discretized version Feynman path integral
describing $D$--dimensional geometries undergoing `quantum fluctuations'.
According to this prescription, the asymptotic functional (\ref{PRasymt}) --to be understood
here as the semiclassical limit of a sum over truly `quantum' fluctuations -- turns out to be
associated with  the simplest 3--dimensional `spacetime', the Euclidean tetrahedron.
The construction of the so--called Ponzano--Regge `state sum' representing the
quantum partition function of simplicial $3$--gravity will not be discussed further
(see {\em e.g.} \cite{MARA2} Sec. 5, for an account).

Several years after Ponzano--Regge paper, a regularized version of
their state sum
--based on representation theory of a `quantum' deformation
of the group $SU(2)$--  was proposed in \cite{TUVI} and shown to be a
well--defined (finite) topological invariant for closed 3--manifolds.
Since then there has been a renewed interest also in the asymptotics (\ref{PRasymt})
both in connection with the study of $3$--manifold geometry (and higher--dimensional
generalizations \cite{CAMA1}) and in addressing `loop quantum gravity' models, see
{\em e.g.} \cite{ROV} and references therein.

\subsection{Quantum automata and topological invariants}

In the past few years there has been a tumultuous activity
aimed at introducing novel conceptual schemes for quantum
computing. The model of quantum simulator
proposed in \cite{MARA1,MARA2} and further discussed in \cite{MARA3,MARA4}
relies on the recoupling theory of $SU(2)$ angular momenta
and can be viewed as a generalization to
arbitrary values of the spin variables of the usual quantum--circuit
model based on `qubits' and Boolean gates \cite{NICH}.
The basic ingredient of such general scheme for universal quantum computing
are indeed encoded into  spin network computational graphs
of the type depicted in
the central portion of Fig. \ref{fig1}). Such pictorial representation makes it
clear that the computational space of the simulator complies with the architecture of
(families of) `automata'\footnote{An automaton in computer science is
a graph whose nodes encode `internal states' while a link between two nodes
represents an admissibile  `transition' between the corresponding states.}.
According with this kind of interpretation, a computational process on
the spin network can be associated with a {\em directed path}, namely
an ordered sequence of vertices and edges starting from an initial quantum
state, say $|s>_{in}$, and ending in some set of final states $\{|s>_{fin}\}$.

In a series of papers \cite{GAMARA1,GAMARA2,GAMARA3,GAMARA4,GAMARA5} families of automata arising from
the $q$--deformed analog of the spin network simulator have been implemented
in order to deal with classes of computationally--hard problems
in geometric topology (topological invariants associated with knots
and with closed $3$--dimensional manifolds)\footnote{A topological invariant is
a quantity --typically a number or a polynomial-- that depends only
on the global topology of the geometrical object and not on its local metric
properties.}.

From the point of view of classical complexity theory,
computing such invariants is `hard', namely could be achieved by
a classical computer only by resorting to an exponential amount of resources.
A computational process which requires an amount of resources that grows
at most polynomially with the size of the computational problem
is referred to as `efficient' ({\em cfr.} \cite{GAMARA5}
and references therein for an account of algorithmic questions
involving braid group and topological invariants ok knots).

In \cite{GAMARA1,GAMARA2}, efficient ({\em i.e.}
running in polynomial time) quantum algorithms for
approximating, within an arbitrarily small range,
$SU(2)_q$--colored polynomial invariants of knots have
been explicitly worked out. In
\cite{GAMARA4} such algorithms have been generalized to deal
with $3$--manifold invariants, while in \cite{GAMARA3}
connections among quantized geometries, topological quantum
field theory and quantum computing
are discussed in detail.

\section{\label{sec:sei}Extensions and applications}

Besides $q$--deformation, for alternative extensions of spin
networks see \cite{125}, where `ternary trees' are introduced
to represent graphically the basic features of `elliptic'
coordinate sets on $S^2$ and $S^3$ ($S^d$ denotes the
standard $d$--sphere embedded in the $(d+1)$--dimensional
Euclidean space) and of the corresponding harmonics.
Interestingly, moving continuously along the edges of the abacus of
Fig. \ref{fig1} can be associated to the variation of the
modulus of elliptic functions \cite{272,273,279}.

In Sec. \ref{sec:quattroB}, we have outlined how the concepts of
reference frame transformations and of alternative angular momentum coupling
schemes in quantum mechanics lead to different representations of
the quantum scattering matrix and provide a powerful guide for the
analysis of atomic and molecular collisions. In particular, we
have exemplified atomic and molecular elastic and inelastic
collisions, but extensions to reactive scattering are most interesting and
extensive applications have been worked out.
Dynamical calculations for the system He+H$_2^+$ \cite{240,239}
and for the benchmark reaction F+H$_2$ \cite{228,259} have been
performed, also including fine--structure and isotopic effects on reactivity.

Also, a new class of entrance and exit channel indices in the scattering
matrix has been worked out.
Through the hyperspherical coordinate formulation referred
to in Sec. \ref{sec:quattroB}, the hyperrradial problem is essentially equivalent to
that of scattering  from anisotropic potentials, and such a
``stereodirected representation'' of the scattering matrix can be used
to derive information about the stereodynamincs of an atom--diatom
reaction. A quantity that can be reconducted to properties
measured in beam experiments on oriented molecules is the
reaction probabilitiy as a function of the {\em steric quantum
number} (see {\em e.g.} \cite{245}).

Since present quantum--mechanical calculations are becoming feasible for
reactive encounters on realistic potential energy surfaces (the
hyperquatization algorithm provides an efficient machinery in this
direction), stereodynamical properties exploiting the
stereodirected representation have been reported, specifically for the
reaction of HF with Li \cite{211,224} and of F with H$_2$ \cite{282}.

Further perspectives concerning the extension of angular momentum theory
to hyperspaces and the use of modern advances in the theory of
orthogonal polynomials of a discrete variable have been
reviewed \cite{236}. Among applications, it is worth mentioning the
possibility of representing polarization parameters by `discrete'
multipole moments \cite{209,307} and potential energy surfaces by
orthogonal discrete basis sets \cite{244}.

The study of asymptotic expansions of $3nj$ symbols for $n>2$, as
well as of different types of asymptotics (in which only a few
variables are large, while the other ones are kept  `quantized')
represents a major challenge not only in the framework of the
formal theory of hypergeometric polynomials and related hierarchies
\cite{KOE98}, but also in view of applications to specific physical
problems arising in connections with all the topics discussed in the
previous Sections \ref{sec:quattro} and \ref{sec:cinque}.
A key example is the crucial occurrence of $9j$ symbols in the many
center problem in quantum chemistry, either in Sturmian orbital or
in momentum space approaches \cite{266,AVERY}.

Finally, for what concerns the issues discussed in Sec. \ref{sec:cinque},
further algorithmic problems regarding  (spin network--type) quantum geometry,
topological quantum field theories in dimension $3$ and associated
$2$--dimensional lattice models (as well as relations among them)
are currently addressed \cite{KAMARA}.

\section{Acknowledgments}

Thanks are due to Robert Littlejohn [Berkeley, California], RogerAnderson [Santa Cruz, California]
and Mario Rasetti [Turin, Italy]. Doctoral fellowships to ACPB by Capes, Brazil, and to CSF by Alban,
EU, is gratefully acknowledged. Support by Italian Agencies ASI and MIUR are also acknowledged.

\end{document}